\documentclass[aps,prd,nofootinbib,preprint]{revtex4}
\usepackage[colorlinks=true, pdfstartview=FitV, linkcolor=blue, citecolor=red, urlcolor=magenta]{hyperref}
\usepackage{graphicx}
\usepackage{latexsym}
\usepackage{amsmath}
\usepackage{amsfonts}
\usepackage{amssymb}
\usepackage{verbatim}
\usepackage{dcolumn}
\usepackage{amssymb}
\usepackage{bm}
\usepackage{float}
\usepackage[english]{babel}
\newcommand{\be}{\begin{equation}}
\newcommand{\ee}{\end{equation}}
\newcommand{\ba}{\begin{align}}
\newcommand{\ea}{\end{align}}

\begin{document}

\newcommand{\JPess}{
\affiliation{Departamento de F\'isica, Universidade Federal da Para\'iba, \\Caixa Postal 5008, 58059-900, Jo\~ao Pessoa, PB, Brazil}
}

\newcommand{\Roma}{\affiliation{Dipartimento di Fisica, Universit\`a di Roma ``La Sapienza", P.le A. Moro 2, 00185 Roma, Italy}}

\title{Rainbow-like Black Hole metric from Loop Quantum Gravity}

\author{Iarley P. Lobo}
\email{iarley\_lobo@fisica.ufpb.br}
\JPess
\author{Michele Ronco}
\email{michele.ronco@roma1.infn.it}
\Roma

\begin{abstract}
The hypersurface deformation algebra consists in a fruitful approach to derive deformed solutions of general relativity based on symmetry considerations with quantum gravity effects, whose linearization has been recently demonstrated to be connected to the DSR program by the $\kappa$-Poincar\'e symmetry. Based on this approach, we analyzed the solution derived for the interior of a black hole and we found similarities with the, so called, rainbow metrics, like a momentum-dependence of the metric functions. Moreover, we derived an effective, time-dependent Planck length and compared different regularization schemes.
\end{abstract}


\maketitle



\section{Introduction}\label{intro}

Despite tenacious and enduring efforts along many years of research, the sought dream of quantizing gravity is still far from being accomplished. Various attempts, which seemed particularly promising at their birth, got stuck with insurmountable obstacles in the form of several formal complexities \cite{oriti,smol,gacLRR,RovelliLRR,alvarez,nicolai1,nicolai2}. In the light of this, a more pragmatic approach to the problem of quantum gravity (QG) consisted in looking for simplified (or, better to say, effective) models able to encode a few characteristics of what we expect to be the theory of QG \cite{gacLRR,smam,liberati1,liberati2,mattingly}. Of course these models could not provide us with the ``final theory" but may capture some key ingredients of QG, optimistically those that may allow us to perform experimental tests needed to guide our intuition as well as the construction of more reliable formal approaches to the problem. Typically, full-fledged QG approaches and more phenomenological models moved along parallel tracks. However, in the last few years some steps to shorten the gap between these two complementary views have been taken. 
\par
Given the complexity and variety of the QG panorama, it is useful and common to divide different approaches in two broad categories: covariant approaches and canonical approaches. The former class is based on the assumption of diffeomorphism invariance and seems to leave no room for quantum deformations of it. On the other hand, the canonical procedure makes the covariance of general relativity (GR)  less evident by construction \cite{dirac,adm,thibook,bojobook,CorichiReyes} and, indeed, symmetries need to be checked directly by means of the calculation of the Poisson brackets between gravitational constraints. Interestingly, such a procedure has been recently proven to allow for modifications of GR covariance that preserve a certain symmetry structure in a deformed sense \cite{covqg1,covqg2}. We feel this could be insightful for the construction of QG models as well as the relations between different models and, hopefully, also for its phenomenological signatures. 
\par
In particular, the approach of canonical loop quantum gravity (LQG) \cite{thibook,bojobook,CorichiReyes}, that counts remarkable accomplishments such as singularity resolution in various cosmological and black-hole scenarios and a meaningful space discretization, faces major difficulties in finding a quantum realization of the Hamiltonian, difficulties which so far remained unsolved \cite{alex,Amb2,Amb3}. Given that, a number of recent analyses \cite{CaitellMielcBarr, AshtLewMarMouThie, bojopail2,bojoHossKag} have tried to circumvent such a problem by constructing canonical effective theories of QG, analogously to what is being done for modified theories of gravity in the study of dark matter or dark energy with several toy models. In this way, one can write modified gravitational constraints (i.e. Hamiltonian density and momenta) which take into account quantum corrections in the form of non-linear modifications of the phase-space variables inspired by the LQG quantization techniques. Remarkably, at least in symmetry-reduced LQG models, it has been shown that the modified constraints still form a closed set of Poisson brackets, which is alternatively deformed with respect to the case of ADM GR \cite{bojopail2,Thiemann2,CuttSakell,BojoBraRey,B1,B2,B3}.  These first studies, performed within the framework of effective LQG, attracted renewed interest in the possibility of QG-induced symmetry deformations and inspired further analyses in other approaches beyond general relativity, namely: the gravity sector of multi-fractional models \cite{calcagni}, a certain class of (minimally) modified theories of gravity in the canonical formulation \cite{liberati1}, and finally canonical noncommutative gravity with $\star$-product deformations of the algebra \cite{moyaldiff}. Thus, this gave additional support to claim that QG may require a deformation of GR covariance. 
\par
Intriguingly, the possibility of symmetry deformations induced by quantum effects is not something new in the QG research but rather a recurring idea which, from time to time, has taken different concrete forms in the literature. Most significantly, it is at the core of the class of models that goes under the name of deformed (or doubly) special relativity (DSR) where the Planck length, the characteristic scale of QG physics, is supposed to play the role of a relativistic invariant scale analogously to the speed of light \cite{dsr1,dsr2,Magueijo:2002am,dsr3,dsr4}. Concretely, such a proposal has been realized in the studies of non-commutative spacetime geometries where, as a consequence of spacetime non-commutativity, the special relativistic symmetries are modified by Planckian corrections and in some cases, most notably in the so-called $\kappa$-Minkowski geometry which is the non-commutative spacetime dual to the $\kappa$-Poincar\'e algebra, $M_P$ actually represents a relativistic invariant quantity \cite{lukRue,majRue}. For the purposes of this work, it is of particular importance the fact that, in the Minkowski limit, the LQG-deformed symmetries 
are consistent with the $\kappa$-Minkowski noncommutative spacetime as shown by one of us in \cite{kminklqg}. 
\par
The main importance of the results on covariance derived in modified canonical models is that they may serve to bridge the gap between LQG and observable low-energy physics. In fact, some studies \cite{mdrlqg1,mdrlqg2} have outlined how modified dispersion relations (MDR), i.e. Planck-scale corrections to the on-shell relation, and a reduction of dimensions at the Planck scale \cite{lqgdim1,lqgdim2} can be derived from the modified brackets of gravitational constraints. Moreover, the analysis of \cite{mrPLB} suggested that the type of modifications introduced in the gravitational constraints affects directly the form of the MDR in such a way that future tests of Planck-scale departures from special relativistic symmetries could hopefully distinguish different theoretical scenarios in the not too distant future. Within the context of deformed covariance, another strategy to extract phenomenology could be the computation of effective metrics from the LQG-deformed constraint equations. Such an approach has already proved its richness in the case of loop quantum cosmology where effective Friedmann-Robertson-Walker (FRW) spacetimes allowed researchers to find robust solutions to  the singularity problem as in bouncing cosmological models. Very recently,  effective line elements for black-hole models have been derived by solving deformed Einstein-like equations implied by the deformed algebra of constraints \cite{Bojowald:2018xxu,BenAchour:2018khr}. This opens the way to the investigation of semi-classical black hole solutions with  LQG corrections. 
\par
In both cases, one has to address the issue of coarse-graining at larger scales (i.e. lower energies) the microscopic texture of the geometry, which at the Planck scale is described non-perturbatively by quantum operators and the associated states on a Hilbert space. It is worth stressing that a satisfactory definition of the (semi-) classical continuum limit has not been accomplished yet by working within the full complexity of the LQG formalism. However, several encouraging results have been obtained in the context of symmetry reduced models. In those cases, the problem of dynamics is greatly simplified and an analytic expression for the scalar constraint can be found. Then, semi-classical states are defined by peaking around classical trajectories and it has been shown that these states exponentially dominate the partition function that sum over geometries \cite{Bianchi:2009ky}. Thus, effective models can be eventually considered in analogy with gauge theories defined on a discrete lattice, whereby the constraint operators are regularized by some lattice parameter identified with (or close to) the Planck length. As a consequence, the continuum limit is automatically obtained once such a regulator is removed.
\par
\par
\par
Here, building on the results of \cite{BenAchour:2018khr}, we show that the LQG modifications of the black hole metric can be written as functions of the total radial momentum, thereby introducing an explicit dependence of the metric on the Poincar\'e charges as proposed in the approach of \textit{rainbow gravity} (RG) \cite{dsr2,Magueijo:2002am,Magueijo:2002xx,Galan:2004st,Ali:2014xqa,Heydarzade:2017rpb}. Such an observation we here put forward to strengthen the lacking synergy between fundamental approaches and phenomenological toy models which can be important in order to both improve our intuition about the formal structures required by the QG theory and, at the same time, conceive experimental tests of potential Planck-scale effects.  
\par
The paper is organized as follows: in section \ref{sec-def} we review some basics properties of the algebra of gravitational constraints (or HDA) and its deformations from LQG corrections, as well as the modified black-hole solutions derived in \cite{BenAchour:2018khr}. In section \ref{sec-rain} we further analyze this map in order to map the effective metric found in \cite{BenAchour:2018khr} and rainbow metrics. In section \ref{sec-disc} we compare our results with the ones previously found in the literature. Then we conclude in section \ref{sec-conc}.


\section{Hypersurface deformation algebra}\label{sec-def}
In the last decade, one of the most interesting results in LQG has been the emergence of non-classical spacetime structures from simplified analyses relying on effective field theory models for QG. These departures from smooth classical spacetime manifolds can be meaningfully traced back to quantum modifications of the so-called hypersurface deformation algebra (HDA). In classical Hamiltonian GR the HDA is given by the following set of Poisson brackets \cite{dirac,adm}

\begin{equation}\label{hda}
\begin{split}
\{D[M^{a}],D[N^{a}]\}=D[\mathcal{L}_{\vec{M}}N^{a}],\\
\{D[N^{a}],H[M]\}=H[\mathcal{L}_{\vec{N}}M],\\
\{H[M],H[N]\}=D[ h^{ab}(M\partial_{b}N-N\partial_{b}M)] \, , 
\end{split}
\end{equation}
which encodes covariance in the Hamiltonian formulation of classical GR. Here $D[M^{a}]$ is the momentum (or spatial) constraint that generates deformations along the 3-dimensional hypersurfaces by an amount $M^a$ (with $a= 1,2,3$), while $H[N]$ is the Hamiltonian (or time) constraint responsible for translations along the normal direction to these hypersurfaces, finally $h^{ab}$ are the components of the inverse three metric. More precisely, given a generic phase-space function $f(h_{ij}, \pi^{ij})$, being $\pi^{ij}$ the gravitational momentum conjugate to the metric, one has that 

\begin{equation}
\delta_{\overrightarrow{M}} f(h_{ij}, \pi^{ij}) = \{ f(h_{ij}) , D[\overrightarrow{M}]\} \, , \quad \delta_{N} f(h_{ij}) = \{ f(h_{ij}, \pi^{ij})  , H[N]\}.
\end{equation}

The search for a quantum version of the gravitational constraints represents the main objective of the approach known as \textit{canonical quantum gravity} and, so far, has not been conclusive. However, in spite of the fact that the full quantum theory is not available (mainly due to the renowned difficulties in the regularization of the Hamiltonian operator), consistency relations implied by the desire to preserve spacetime symmetries can be used to identify an effective formulation of LQG where a consistent set of closed Poisson brackets can be found by introducing restricted and simplified correction functions into the Hamiltonian, which are inspired by the LQG quantization technique (see e.g. \cite{Barrau} and references therein).
\par
Therefore, the first step consists in correcting the classical scalar and diffeomorphism constraints with possible modifications motivated by LQG. There is a certain degree of arbitrariness in the specific choice of these correction functions, greater than what is commonly acknowledged, and this will be discussed in some detail later on (see also \cite{mrPLB}). Nonetheless, we can fairly divide them into two broad classes: \textit{inverse triad} and \textit{holonomy} corrections. We will here consider only the latter type of quantum (or, better to say, semi-classical) contributions. They can be motivated by the fact that holonomies of the Ashtekar connections are the basic LQG variables and, at the effective level, can be taken into account by replacing the mean connection by a periodic function  in the Hamiltonian constraint. The former are the inverse-triad corrections that come from terms in the Hamiltonian constraint which cannot be quantized directly but only after being re-expressed as a Poisson bracket, a procedure which is usually referred to as the Thiemann-trick for the quantization of $H[N]$ \cite{thieTr}. However, as already said, they will not be
contemplated here. One then works with (modified) classical phase-space functionals which can be understood as the result of the evaluation of distribution valued operator over an orthonormal basis in terms of spin network states that span the Hilbert space.  These quantum corrections may or may not spoil the symmetry of the classical theory under diffeomorphisms. Indeed, one has to prove that the quantum-corrected constraints form a closed algebra thereby eliminating the same number of spurious degrees of freedom as in the classical theory given their role of generators of gauge transformations. This poses the issue of anomaly freedom, which is the focus of the so-called \textit{deformed-algebra approach} to (effective) LQG \cite{bojopail2,Thiemann2,CuttSakell,BojoBraRey,B1,B2,B3,bra1,bra2}. The goal consists in introducing these effective quantum corrections into the classical gravitational constraints and, then compute the Poisson brackets between them in order to check the compatibility with the symmetry under diffeomorphism. A closure of the HDA despite the presence of holonomy corrections would imply that symmetries are preserved, and it could be regarded as a strong hint that LQG is not anomalous. On the other hand, any kind of modifications to the brackets \eqref{hda} could signal that diffeomorphism transformations are deformed due to ``quantum" effects. 
\par
In particular, one starts from polymerizing the angular extrinsic curvature component:

\begin{equation}
\label{Rholocorr}
K_{\phi}^2 \rightarrow h(K_{\phi}) =  \frac{[\sin(\rho K_{\phi})]^2}{\rho^2}\,,
\end{equation}
where $\rho$ is related to some scale, usually $\ell_P$, as suggested, for instance, by the discrete spectrum of the area operator ($\rho$ is proportional to the square root of the minimum eigenvalue, or the `area gap' from LQG) or on the size of the loop considered for the definition of holonomies. Clearly, the classical regime is recovered in the limit $\rho \longrightarrow 0$.\footnote{ The fact that zero does not belong to the spectrum of the area operator in LQG is precisely the input from the full theory which gives a nontrivial quantum geometrical effect.} The above substitution \eqref{Rholocorr} can be justified as follows. In the quantum theory there is no well-defined operator corresponding to the Ashtekar-Barbero  connection $A^{i}_{a}$ on the LQG kinematical Hilbert space. Instead, in the loop representation, a well-defined object is the holonomy operator which are defined as parallel transport of the connection

\begin{equation}
\label{holo}
h_{\alpha}(A) = \mathcal{P}\exp(\int_{\alpha}\dot{e}^{a}A^{i}_{a}\tau_{i})\,,
\end{equation}
where $\mathcal{P}$ is the path-ordering operator and $\dot{e}^{a}$ is the three vector tangent to the curve $\alpha$. For our analysis are of particular interest the holonomies of connections along homogeneous directions, which simplify as \cite{bojoHossKag}

\begin{equation}
\label{homoholo}
h_{j}(A) = \exp(\mu A \tau_{j}) = \cos(\mu A) \mathbb{I}+\sin(\mu A)\sigma_{j}
\end{equation}
and do not require a spatial integration since they transform as scalars. In fact, so far  one knows only how to implement (local) holonomy corrections for connections along homogeneous directions (for a negative result concerning implementation of nonlocal (extended) holonomy corrections in spherical symmetry see \cite{SSnonlocalhol}). In our case, this is given by $\gamma K_{\phi}$ ($= A_{\phi}\cos \alpha$):

\begin{eqnarray}
\label{angholo}
h_{\phi} ( r,\mu) &=& \exp(\mu A_{\phi}\cos\alpha \Lambda^{A}_{\phi})\nonumber\\
 &=& \cos(\mu \gamma K_{\phi}) \mathbb{I}+\sin(\mu \gamma K_{\phi}) \Lambda
\end{eqnarray}

In order to see how the replacement \eqref{Rholocorr} is implied by Eq. \eqref{angholo} one must take into account that the scalar constraint is quantized by  utilizing the Thiemann trick $\sqrt{E^{r}}  \propto \{ K_{\phi}, V \}$ (where $V$ is the volume), whose quantum version contains the commutator $h_{\phi}[h^{-1}_{\phi},\widehat{V}] = h_{\phi}h^{-1}_{\phi}\widehat{V}- \widehat{V}h^{-1}_{\phi}\widehat{V}h_{\phi}$. (This is equivalent to regularizing the curvature of the connection by holonomies, with the minimum area being the `area gap' from LQG.) Using Eq. \eqref{angholo} one can easily see that products of holonomies are given by cosine and sine functions of $K_{\phi}$. Finally, it turns out that the resulting quantum or `effective' (since we are going to ignore operator ordering issues by working in a semi-classical setting, which are not crucial to our goals) scalar constraint could be obtained simply making the replacement of Eq. \eqref{Rholocorr}. This justifies the following form of the effective Hamiltonian constraint $H^Q$

\begin{equation}
\label{quantsc}
H^{Q}[N] = -\frac{1}{2G}\int_{B} dr N \left[ \frac{[\sin(K_{\phi}\rho)]^2}{\rho^{2}}E^{\phi}+ 2K_{r}\frac{\sin(K_{\phi}\rho)}{\rho}E^{r} +(1-\Gamma_{\phi}^{2})E^{\phi}+2\Gamma_{\phi}^{'}E^{r}\right]\, . 
\end{equation}

On the other hand the diffeomorphism constraint remains undeformed since spatial diffeomorphism invariance translates into vertex-position independence in LQG, which is implemented directly at the kinematical level by unitary operators generating finite transformations\footnote{In fact, there is no well-defined infinitesimal quantum diffeomorphism constraint in LQG for the basis spin network states. Some progress in constructing it has been achieved in \cite{Varadarajan}.}.\\
As aforementioned, the crucial point of the deformed algebra approach is to ensure that the resulting algebra of constraints remains consistent, so that Poisson brackets between quantum corrected constraints are proportional to a quantum corrected constraint. Such a procedure has to be performed ``off-shell", i.e. before  the quantum corrected equations have been solved. In the case of gravity, this is the only way to guarantee that the quantum theory is fully consistent. With a rather straightforward but lengthy calculation one can show that the gravitational constraints with LQG corrections close the algebra non-pertubatively.  Particularly remarkable is the fact that, at least for symmetry reduced cases, there is a unique solution to the anomaly freedom problem. In fact, the full deformed-HDA is given by \cite{bojopail2,Thiemann2,CuttSakell,BojoBraRey,B1}

\begin{equation}\label{qhda}
\begin{split}
\{D[M^{a}],D[N^{a}]\}=D[\mathcal{L}_{\vec{M}}N^{a}],\\
\{D[N^{a}],H^{Q}[M]\}=H^{Q}[\mathcal{L}_{\vec{N}}M],\\
\{H^{Q}[M],H^{Q}[N]\}=D[\beta h^{ab}(M\partial_{b}N-N\partial_{b}M)] \, .
\end{split}
\end{equation}

Thus, these modifications amount to a deformation of the brackets closed by the gravitational constraints which generate space and time gauge transformations. Specifically, only the Poisson bracket involving two Hamiltionain constraints is modified by the presence of a deformation function  that depends on the phase space variables, i.e. $\beta = \beta(h_{ij}, \pi^{ij})$ (or, equally, $\beta = \beta (A^a_i, E^j_b)$), whose particular form depends on the specific holonomy corrections considered as well as on the symmetry reductions implemented and so forth.
\par
The angular component of the extrinsic curvature $K_\phi$ can be consistently quantized and produces the above result. To see that, we have to briefly introduce the spherically-symmetric reduction of Hamiltonian gravity in Ashtekar-Barbero variables (see e.g.\cite{olmedo}) in the presence of LQG deformations. In this case the ADM foliation ~\cite{adm} allows a decomposition of the spacetime manifold as $\mathcal{M} = \mathbb{R}\times \Sigma = \mathcal{M}_{1+1}\times S^2$, where $\mathcal{M}_{1+1}$ is a 2-dimensional manifold spanned by $(t,r)$ and $S^2$ stands for the 2-sphere. Given that, the line element reads
\begin{equation}
\label{metric}
ds^2 = -N^2dt^2+h_{rr}(dr+N^r dt)^2 + h_{\theta\theta}[d\theta^2+(\sin(\theta))^2d\phi^2]\,,
\end{equation}

where the shift vector is purely radial, {\it i.e.} $N^i = (N^r,0,0)$, due to spherical symmetry, and, consequently, we are left only with radial diffeomorphisms generated by $D[N^r] = \int dr N^r \mathcal{H}_r$ (where $\mathcal{H}_r$ is the only non-vanishing component of the momentum density) and, time transformations, generated by $H[N] = \int dr N \mathcal{H}$ (where $\mathcal{H}$ is the Hamiltonian density). The components of the spatial metric $(h_{rr},h_{\theta\theta})$ can be written in terms of rotationally invariant densitized triads which are given by:

\begin{equation}\label{triads}
E = E^{a}_{i}\tau^{i}\frac{\partial}{\partial x^{a}}=E^{r}(r)\tau_{3}\sin\theta\frac{\partial}{\partial r}+E^{\phi}(r)\tau_{1}\sin\theta\frac{\partial}{\partial\theta}+E^{\phi}(r)\tau_{2}\frac{\partial}{\partial\phi}\,,
\end{equation}

where $\tau_{j}=-\frac{1}{2}i\sigma_{j}$ represent \textit{SU}(2) generators. The densitized triads are canonically conjugate to the extrinsic curvature components, which, in presence of spherical symmetry, are conveniently
described as follows

\begin{equation}
K=K^{i}_{a}\tau_{i}dx^{a}=K_{r}(r)\tau_{3}dr+K_{\phi}(r)\tau_{1}d\theta+K_{\phi}(r)\tau_{2}\sin\theta d\phi\,.
\end{equation}

As a result, the components of the three metric are

\begin{equation}
h_{\theta\theta} = E^r(r) \, , \quad h_{rr} = \frac{(E^\phi (r) )^2}{E^r (r)} \, .
\end{equation}

At this point, one can show that the bracket $\{H^Q[N], H^Q[M] \}$ in Eq. \eqref{qhda} reads

\begin{equation}
\{H^Q[N], H^Q[M] \} = D[\beta(\rho K_\phi) \frac{E^r}{(E^\phi)^2} (N\partial_r M - M \partial_r N)] \, , 
\end{equation}

where $\beta$ is related to the second derivative of the holonomy-correction function, i.e. $\beta = h^{''}/2$.  In particular, for the simplest case including only local holonomy corrections as in Eq. \eqref{Rholocorr}  (see also \cite{ed,fra} for a detailed construction and the related discussion), with $\gamma \in \mathbb{R}$ and $j=1/2$, the deformation $\beta$ takes the form

\begin{equation}
h = \frac{[\sin(\rho K_\phi)]^2}{\rho^2} \, \quad \Longrightarrow \, \quad \beta=\cos(2\rho K_{\phi}) \, .
\end{equation}

However, more complicated expressions are possible and will be discussed in the next section. As shown explicitly in \cite{BenAchour:2018khr}, given the modified HDA, one can then obtain Einstein-like equations of motion with LQG corrections from 

\begin{equation}
\dot{F} = \{ F, H^Q[N] + D[M^r]\} \, , 
\end{equation}
with $F = (E^r, E^\phi, K_\phi, K_r)$. For instance, the equations of motion for the two independent triads, the extrinsic curvature $K_{\phi}$ and the Hamiltonian constraint (which can be used to find $K_r$) are

\begin{equation}
\begin{split}
\dot{E}^r =  N\sqrt{E^r}h'(K_\phi) +M^r \partial_r E^r \, ,   \\
\dot{E}^\phi = \frac{N}{2}\left( \sqrt{E^r}K_r    h^{''}(K_\phi) + \frac{E^\phi}{\sqrt{E^r}}h^{'}(K_\phi)\right) + \partial_r (M^r E^\phi)  \, ,	\\
\dot{K}_{\phi}=-\frac{N}{2\sqrt{E^r}}[1+f(K_{\phi})]\, ,	\\
h'(K_{\phi})E^rK_r+(1+h(K_{\phi}))E^{\phi}=0\, .
\end{split}
\end{equation}

In Ref.\cite{BenAchour:2018khr}, the above LQG-corrected Einstein equations have been solved explicitly for the interior of a static black hole. The solutions for the triads read 

\begin{equation}
E^r = t^2 \, , \quad E^\phi = \frac{r_S}{2} \frac{h^{'}(K_\phi)}{1 + h(K_\phi)} \, ,
\end{equation}
and the of the extrinsic curvature $K_{\phi}$ is
\be\label{solK}
h(K_{\phi})=\frac{r_s}{t}-1\, .
\ee
where $r_S$ is the Schwarzschild radius.\footnote{We omit the solution for $K_r$, because it will not be used in our analysis.} Finally, the LQG-modified line element is

\begin{equation}\label{lqg-line-element}
ds^2 = -\frac{1}{F(t)}dt^2 + F(t) dr^2 + t^2 d \Omega^2 \, , 
\end{equation}
with 

\begin{equation}\label{F1}
F(t) = \left( 2 \frac{d h^{-1}}{d x}\Big |_{x=\frac{r_S}{t}-1}\right)^{-2} \, .
\end{equation}


\section{Effective rainbow metric}\label{sec-rain}
In general, one has as the solution a deformed metric that depends on the spacetime coordinates and on the deformation parameter $\rho$. However, recently the deformation function $h(K_{\phi})$ gained a different role. It was shown that such function, in fact, deforms the Lorentz algebra of the spacetime found in the flat version of the HDA described above, see \cite{kminklqg} and references therein (see also  \cite{mdrlqg1} for a different analysis leading to similar outcomes, i.e. deformed Poincar\'e symmetries in the Minkowski limit of \eqref{qhda}).
\par
For our purposes, it is of pivotal importance to find a way to write $\beta$ in terms of symmetry generators (see also \cite{kminklqg,mdrlqg1,mdrlqg2}), and to this end, it is valuable to notice that observables of the Brown-York momentum~\cite{BY},

\begin{equation}\label{bymom}
P=2\int_{\partial\Sigma} d^{2}z\upsilon_{b}(n_{a}\pi^{ab}-\overline{n}_{a}\overline{\pi}^{ab}) \, ,
\end{equation}
can be identified by extrinsic curvature components. In Eq.\eqref{bymom}, we have that $\upsilon_{a}=\partial/\partial x^{a}$, $n_{a}$ is the co-normal of the boundary of the spatial region $\Sigma$, and $\pi^{ab}$ plays the role of the gravitational momentum (while the over barred symbols in the above equation are the same functions but evaluated at the boundary). From this, it is possible to establish that the radial Brown-York momentum $P_{r}$ is related to the extrinsic curvature component $K_{\phi}$ in the following way (see, e.g., \cite{mdrlqg2})

\begin{equation}\label{bymomf}
P_{r}=-\frac{K_{\phi}}{\sqrt{|E^{r}|}}.
\end{equation}

The flat case was discussed in \cite{kminklqg} in the context of DSR symmetries. In that case, since $E^r$ is a constant, it was possible to set the parameter $\rho\propto |E^r|^{-1/2}$. Which allows to relate the deformation function $\beta$ to the generator of radial translations $P_r$

\begin{equation}\label{funb}
\beta = \cos(\lambda P_{r}) \, , 
\end{equation}
where $\lambda$ is a parameter of the order of the Planck length ($\lambda \sim \ell_P \sim 1/ M_P$).\footnote{Keep in mind that its exact value also depends on quantization ambiguities \cite{mrPLB}.} Such identification allowed the authors to derive deformed relations for the symmetry generators of the flat spacetime.
\par
That is, on one side this approach traces a map between DSR and the Minkowski limit of the HDA from the point of view of deformed symmetries. And on the other side it was recently found an exact solution of the field equations derived from a HDA for the curved case of the black hole interior. In principle, these two approaches are independent, i.e., there is no local DSR description of the symmetries of the deformed metric yet, nor a metric description that emerges from this DSR proposal.
\par
A metric description that is able to encode these aspects of the formalism is still unknown. In this paper we aim to contribute to this subject by describing the curved effective metric in the light of the discovered relation between HDA and DSR. In fact, a relevant approach to the metric description inspired by the DSR scenario conjectures that the spacetime metric determined by an observer by measurements done with an energetic particle depends on the particle's energy as measured by that observer. The deformed relativistic metric description should be given in terms of a rainbow metric \cite{Magueijo:2002xx}. Therefore, we want to study whether the intuition of rainbow gravity finds support in this recently found effective curved metric from HDA, when one uses the prescription that relates HDA and DSR.


\subsection{Rainbow gravity}
In this subsection we review the main aspects of the standard rainbow gravity as proposed in \cite{Magueijo:2002xx}. In this case, consider a MDR of the type\footnote{We are considering $c=\hbar=1$, which implies in having the Planck length as the inverse of the Planck energy $\ell_P=E_P^{-1}$.}
\be
m^2=E^2f_1^2(\ell_P E)-p^2f_2^2(\ell_P E),
\ee
that can be represented by a simple norm $m^2=\eta^{\mu\nu}U[p]_{\mu}U[p]_{\nu}$, where $U$ is the map in momentum space
\be
U[p]_{\mu}=\left(U[p]_0,U[p]_i\right)=\left(Ef_1(\ell_P E),p_if_2(\ell_P E)\right) \, ,
\ee
where greek indices, like $(\mu,\nu)$, run from $0,...,4$ and latin indices, like $(i,j)$, run from $1,...,3$.
\par
The idea is to write this dispersion relation with an energy-dependent metric $\tilde{\eta}^{\mu\nu}(\ell_P E)$, such that $\eta^{\mu\nu}U[p]_{\mu}U[p]_{\nu}=\tilde{\eta}^{\mu\nu}(\ell_P E)p_{\mu}p_{\nu}$, which could also be generalized for a curved spacetime. A simple way for achieving this consists in transforming the orthonormal frame as $\tilde{e}_A^{\ \mu}=(f_1(\ell_P E)e_0^{\ \mu}\, ,f_2(\ell_P E)e_I^{\ \mu})$, such that
\be
\eta^{\mu\nu}U[p]_{\mu}U[p]_{\nu}=\eta^{AB}\tilde{e}_A^{\ \mu}\tilde{e}_B^{\ \nu}p_{\mu}p_{\nu},
\ee
which defines an energy-dependent metric $\tilde{\eta}^{\mu\nu}(\ell_PE)=\eta^{AB}\tilde{e}_A^{\ \mu}\tilde{e}_B^{\ \nu}$. Here, indices like $(A,B)$ run from $0,...,4$ and ones like $(I,J)$ run from $1,...,3$.
\par
This construction can be directly generalized to curved vielbeins, in fact if one uses the same definition above, it is possible to construct a metric
\be
\tilde{g}^{\mu\nu}(\ell_P E)=\eta^{AB}\tilde{e}_A^{\ \mu}\tilde{e}_B^{\ \nu}, 
\ee
whose inverse is given by
\be
\tilde{g}_{\mu\nu}(\ell_P E)=\eta_{AB}\tilde{e}^A_{\ \mu}\tilde{e}^B_{\ \nu}, 
\ee
where $\tilde{e}^A_{\ \mu}=\left((f_1(\ell_P E))^{-1}e^0_{\ \mu}\, ,(f_2(\ell_P E))^{-1}e^I_{\ \mu}\right)$: this is a {\it rainbow metric}.\footnote{Energy-momentum dependent metrics, like in curved momentum space have been originally considered in \cite{born}.} This way, one can use this kind of metric as an input into the Einstein equations as an ansatz for the so called {\it rainbow gravity}. For instance, a known solution \cite{Magueijo:2002xx} of the Einstein equation for the static and spherically symmetric case is the metric
\be\label{rainbow0}
ds^2=-\frac{1-2M/r}{\left(f_1(\ell_P E)\right)^2}dt^2+\frac{(1-2M/r)^{-1}}{\left(f_2(\ell_P E)\right)^2}dr^2+\frac{r^2}{\left(f_2(\ell_P E)\right)^2}d\Omega^2.
\ee
So, this is a deformation of the Schwarzschild line element by functions that depend on the energy of the particles that probe such spacetime. And since this is a static spacetime, the energy of a test particle is, in fact, a conserved quantity and corresponds to the generator of time translations in this manifold. Thus, implying that the Schwarzschild metric is being essentially deformed by the time-translation generator.
\par
When crossing the horizon, the roles of the radial and the time coordinates change. Such modification takes the metric from a static configuration to a purely time-dependent tensor, which also implies that the energy acquires the role of the conserved radial momentum, i.e., the generator of radial translations. In fact, the metric assumes the form
\be\label{rainbow1}
ds^2=-\frac{(2M/t-1)^{-1}}{\left(f_2(\ell_P P_r)\right)^2}dt^2+\frac{2M/t-1}{\left(f_1(\ell_P P_r)\right)^2}dr^2+\frac{t^2}{\left(f_2(\ell_P P_r)\right)^2}d\Omega^2.
\ee

In the next section, we are going to compare this rainbow metric inside the event horizon of a black hole with the one found from the HDA.

\subsection{Momentum-dependent metric}
Using Eq.(\ref{solK}) we can write (\ref{F1}) as   
\be\label{f1}
F(t)=\left[2\frac{dK_{\phi}(r_s/t-1)}{d\left(r_s/t-1\right)}\right]^{-2}.
\ee
However, if we define $G(r_s/t-1)\doteq dK_{\phi}(r_s/t-1)/d(r_s/t-1)$, which using (\ref{solK}), allows us to define the $K_{\phi}$-dependent function $\widehat{G}(K_{\phi})\doteq G\circ h(K_{\phi})$.
\par
Recalling the relation between the extrinsic curvature and the radial momentum (\ref{bymomf}), we are able to define a metric that presents $P_r$-dependent corrections. It should be stressed that, in this case, $P_r$ corresponds to the quasi-local radial gravitational momentum, which means that it presents the information of the test particle in this spacetime (as described in \cite{kminklqg,Bojowald:2012ux} for deforming the Poincar\'e symmetry, where there is no gravitational field) and of the gravitational interaction (which was absent in the flat case).
\par
To illustrate this construction, let us consider some examples.


\subsubsection{First case}\label{1ex}
The most natural choice to begin our analysis is the one exemplified in Ref.\cite{BenAchour:2018khr}. In this case
\be
h(K_{\phi})=\frac{[\sin(\rho K_{\phi})]^2}{\rho^2}=\frac{r_s}{t}-1
\ee
with
\be
\beta(K_{\phi})=h''(K_{\phi})/2=\cos(2\rho K_{\phi}).
\ee
From Eq.(\ref{f1}), we can derive the coordinate dependence of the metric function $F(t)$ that was found in \cite{BenAchour:2018khr} 
\be
F(t)=\left(\frac{r_s}{t}-1\right)\left[1-\rho^2\left(\frac{r_s}{t}-1\right)\right].
\ee

However in order to analyze this effective metric in the light of rainbow gravity, we propose to take a step back and realize that, in fact, the Schwarzschild metric gets deformed due to the parameter $\rho$ and that such deformation is proportional to $r_s/t-1$, which on the other hand, equals the function $h(K_\phi)$ (by Eq.(\ref{solK})), which in turn is related to the radial momentum $P_r$ by (\ref{bymomf}).

Combining these expressions we have
\be
F(t,P_r)=\left(\frac{r_s}{t}-1\right)[\cos\left(\rho t\, P_r\right)]^2,
\ee
which implies in a rainbow-like metric:
\be
ds^2=-\left(\frac{r_s}{t}-1\right)^{-1}[\cos\left(\rho t\, P_r\right)]^{-2}\, dt^2+\left(\frac{r_s}{t}-1\right)[\cos\left(\rho t\, P_r\right)]^2\, dr^2+t^2\, d\Omega^2.
\ee
\noindent
In this case, the second horizon occurs in the phase space for $\rho t\, P_r=(2n+1)\pi/2$, which corresponds to $t_h=\rho^2r_s/(1+\rho^2)$. In fact, this metric presents the same Penrose diagram, as pointed out in \cite{BenAchour:2018khr}. Also, according to \cite{BenAchour:2018khr,Bojowald:2018xxu}, due to the deformation of the Hamiltonian constraint in Eq.(\ref{qhda}), the time reparametrization of the theory needs to be modified, leading to a rescaling of the lapse function $N$ in Eq.(\ref{metric}) as
\be
N\rightarrow \beta(K_{\phi})N=\cos(2\rho t P_r)N,
\ee
which leads to an Euclideanization of the metric for $\rho t P_r=(2n+1)\pi/4$. For details, see \cite{Bojowald:2018xxu}.
\par
This rainbow metric presents some differences with respect to the usual approach presented before (\ref{rainbow1}). For instance, there is no rainbow function in the angular sector of the line element; this rainbow metric presents contributions from the single particle momentum and from the gravity sector; and the momentum $P_r$ is multiplied by $\rho t$, instead of the usual Planck length $\ell_P$. Such features will be repeated in the next examples. 
\par
Indeed, we warned the reader that holonomy corrections can be implemented in different ways. Specifically, the polymerization function (i.e.  $K \, \mapsto \, f(K)$) depends on some choices we can make such as: the value of the Barbero-Immirzi parameter, the internal gauge group, and finally the spin representation of the group. How these choices affect the symmetry deformation in Eq. \eqref{qhda} and, perhaps, lead to different phenomenological predictions for the form of the MDR has been recently discussed in \cite{mrPLB}. Here, following that line of reasoning, we shall briefly discuss how, as the reader could easily expect, these formal ambiguities affect the shape of these effective rainbow metrics too. 

\subsubsection{Second case}\label{2ex}
A second rather natural choice is represented by the complex Ashtekar variables that, once we turn to the associated effective quantum corrections, gives rise to a similar deformation function through a sort of ``Wick rotation" $\rho\mapsto i\rho$ (see Ref.\cite{mdrlqg2}) of the standard $SU(2)$ polymerization function \eqref{Rholocorr}, i.e. 
\be
h(K_{\phi})=\rho^{-2}[\sinh(\rho K_{\phi})]^2,
\ee
producing the deformed rainbow metric
\be
ds^2=-\left(\frac{r_s}{t}-1\right)^{-1}[\cosh\left(\rho t\, P_r\right)]^{-2}\, dt^2+\left(\frac{r_s}{t}-1\right)[\cosh\left(\rho t\, P_r\right)]^2\, dr^2+t^2\, d\Omega^2.
\ee
In this case, since the hyperbolic cosine is never null, there is just the usual horizon for this black hole. 
\subsubsection{Third case}\label{3ex}
Although complex connection formulations of LQG are receiving restored attention in the recent literature, it is well known that they also raise major difficulties (for instance in the analysis of the observables of the theory, which needs to be real valued operators) nobody has been able to fully and satisfactorily account for. Partial progress is given by the proposal of an ``analytic continuation" procedure (see e.g. \cite{su11}) which has the advantage of preserving the reality of the spectrum of the area operator. We redirect the interested reader to \cite{su11}. From the papers \cite{mdrlqg2,Achour:2014rja} we have
\be
h(K_{\phi})=-\frac{[\sinh(\rho K_{\phi})]^2}{\rho^2}\frac{3}{s(s^2+1)\sinh(\theta_{\phi})}\frac{\partial}{\partial \theta_{\phi}}\left(\frac{\sin(s\theta_{\phi})}{\sinh(\theta_{\phi})}\right),
\ee
where
\be
\sinh\left(\frac{\theta_{\phi}}{2}\right)=\left[\sinh\left(\frac{\rho K_{\phi}}{2}\right)\right]^2.
\ee
Leading to
\begin{align}
ds^2\approx-\left(\frac{r_s}{t}-1\right)^{-1}\left[1+(\rho t)^2\, P_r^2-\frac{(3s^2+4)}{24}(\rho t)^4\, P_r^4\right]^{-1}\, dt^2\\
+\left(\frac{r_s}{t}-1\right)\left[1+(\rho t)^2\, P_r^2-\frac{(3s^2+4)}{24}(\rho t)^4\, P_r^4\right]\, dr^2+t^2\, d\Omega^2.\nonumber
\end{align}
Until the first order, this result coincides with the second case of Section \ref{2ex}, which is coherent with results reported in \cite{mdrlqg2} in the flat case. We consider just the second order deformation due to the complexity of this deformation function.
\subsubsection{Fourth case}\label{4ex} 
Another possibility is represented by higher spin representations of the internal $SU(2)$ group. For instance, in this quantization approach to effective LQG, from \cite{mrPLB}, i.e., $j=1$ HR (holonomy regularization) scheme for regularization, one has:
\be
\beta(K_{\phi})=[\cos(\rho K_{\phi})]^3-[\sin(\rho K_{\phi})]^4-\frac{7}{4}\sin(\rho K_{\phi})\sin(2\rho K_{\phi})+\frac{3}{4}[\sin(2\rho K_{\phi})]^2,
\ee
where $\beta(K_{\phi})=h''(K_{\phi})/2$. Then
\begin{align}
ds^2\approx-\left(\frac{r_s}{t}-1\right)^{-1}\left[1-(\rho t)^2\, P_r^2-\frac{7}{24}(\rho t)^4\, P_r^4\right]^{-1}\, dt^2\\
+\left(\frac{r_s}{t}-1\right)\left[1-(\rho t)^2\, P_r^2-\frac{7}{24}(\rho t)^4\, P_r^4\right]\, dr^2+t^2\, d\Omega^2.\nonumber
\end{align}
We are also considering just a second order approximation, which is sufficient for our discussions.
\subsubsection{Fifth case}\label{5ex}
Now, we consider the case $j=1$, but in the CR (connection regularization) scheme. In this case, following \cite{mrPLB}, we have
\be
\beta(K_{\phi})=[\cos(\rho K_{\phi})]^4-[\sin(\rho K_{\phi})]^4-\frac{3}{2}[\sin(2\rho K_{\phi})]^2.
\ee
Following the same procedures of the previous cases, we are led to the line element:
\begin{align}
ds^2\approx -\left(\frac{r_s}{t}-1\right)^{-1}\left[1-4\, (\rho t)^2\, P_r^2+\frac{16}{3}(\rho t)^4\, P_r^4\right]^{-1}\, dt^2\\
+\left(\frac{r_s}{t}-1\right)\left[1-4\, (\rho t)^2\, P_r^2+\frac{16}{3}(\rho t)^4\, P_r^4\right]\, dr^2+t^2\, d\Omega^2.\nonumber
\end{align}
\section{Comparison with previous definitions}\label{sec-disc}
Now that we have discussed these three cases motivated by three different ways to introduce LQG-inspired corrections into Hamiltonian GR, we recognize that, in general, we obtain metric deformations of the type
\be\label{rainbow-ansatz}
F(t)\approx\left(\frac{r_s}{t}-1\right)[1+\xi_1(\rho t)^2\, P_r^2+\xi_2(\rho t)^4\, P_r^4+\mathcal{O}(\rho t)^6],
\ee
where $\xi_i$ are real numbers. Obviously, the first and second cases could be exactly solved, but they also match this form by performing a Taylor expansion. Therefore we can characterize each of the solutions derived from the HDA by the parameters $\xi_i$ as can be seen in the table (\ref{table1}) for different $j$-representations
\begin{table}[H]
\centering
\caption{$j$-representations and their $\xi_i$-parameters.} 
\begin{tabular}{ccc}
\hline\hline
$j$ & $\xi_1$ & $\xi_2$ \\
$1/2$ & $-1$ & $1/3$ \\ 
$\sim i/2$ & $1$ & $1/3$ \\ 
$\frac{1}{2}(-1+is)$ & $1$ & $-\frac{3s^2~+~4}{24}$ \vspace{3pt}\\ 
$1$ (HR) & $-1$ & $-\frac{7}{24}$ \vspace{3pt}
\\ 
$1$ (CR) & $-4$ & $\frac{16}{3}$ \\ 
\hline\hline
\end{tabular}\label{table1}
\end{table}

The cases analyzed so far do not present deformations as odd functions, therefore terms with odd powers of $t P_r$ cannot appear, i.e., the first order correction appears quadratically, the second order correction appears in 4-th power and so on. This could be the consequence of some symmetry principle underlying the LQG construction that for instance would preserve the parity of the MDR under the transformation $P_r \, \mapsto \,  - P_r$. 
\par
Originally, rainbow gravity was introduced by the so called rainbow functions of the particle's energy $f_{1,2}(\ell_P E)$, which in the case of the Schwarzschild metric inside the black hole reads as (\ref{rainbow0}). There are some fundamental differences with respect to our case. 
\begin{itemize}
\item The usual rainbow function $f_2$ deforms the angular sector of the metric, i.e., the line element of the unit sphere ${\mathbb S}^2$ is momentum-dependent in usual rainbow gravity;
\item Our deforming function $F(t,P_r)$ depends on $\rho t\, P_r$ instead of the usual $\ell_P P_r$;
\item The momentum $P_r$ consists in the momentum of the single test particle and the momentum of the gravitational field.
\end{itemize}
The last two points deserve a further discussion. Regarding the second point, we are lead to speculate whether the rainbow metric inspired by the HDA is deformed by an effective Planck length given by\footnote{In the present case the coordinate time must satisfy $t<r_s$.}
\be
\ell_P^{\text{eff}}=\rho t.
\ee 
\par
This is being generated by the presence of a deformation function on the brackets (\ref{qhda}). Therefore, a possible direction that we could investigate consists in searching for a representation of scalar-tensor theories in rainbow gravity, where the Newton's constant is a scalar field, which would induce a variable Planck length, comparable to what we found the present paper. 
\par
As a matter of fact, if fundamental constants like $\hbar$, $G$ and $c$ are functions of spacetime coordinates this behavior could be explained as long as
\be
\ell_P(t)=\sqrt{\frac{\hbar(t)\, G(t)}{c^3(t)}}=\rho t.
\ee

This is an important difference with respect to previous approaches that build bridges between energy-momentum-dependent metrics and quantum gravity, since for the first time we see a deformation ``parameter" that is coordinate-dependent in this particular context. These possible phenomenological possibilities deserve further investigation. 
\par
Such dependence relies on the relation between the radial momentum and the extrinsic curvature given by Eq.(\ref{bymomf}). For the flat case, there is no coordinate dependence, since the triad $E^r$ is constant, which explains why this feature did not appear in previous analysis of the HDA and DSR, like \cite{kminklqg}. In that case, the term $\rho t$ is replaced by a dimensionful parameter $\lambda$ of the order of the Planck length.
\par
In this regard, we notice that the possibility of a scale-dependence of the characteristic regime at which we should expect QG effects to be relevant is not something new in the literature. Indeed, some kind of running of Planck-scale physics is at the cornerstones of many approaches to the QG problem. Among them we can count causal dynamical triangulation \cite{cdt}, asymptotic safety \cite{as}, and multi-fractional geometries \cite{calcagni2}. In particular, working within this latter approach, one of us \cite{mrPLB2,long} found that the multi-fractional scale (i.e. the ultraviolet scale at which the spacetime dimension changes, as it happens by construction in multi-fractional geometries), $\ell_*$, is related to the scale of the observation at which the measurement is being performed, $s$, i.e. $\ell_* = \ell^2_P/s$ . Within a completely different scenario and framework, we here obtained a similar outcome.  Such an interesting suggestion could be worth exploring elsewhere. 
\par
Also the third point, by itself, deserves a deeper investigation about whether it is possible to uncouple the momenta contributions coming from the gravity and test particle sectors, in order to approximate this new effective metric to the usual one from rainbow gravity, probably in similarity to was done in \cite{Olmo:2011sw} in the context of Palatini $f(R,Q)$ gravity (where $R$ is the usual Ricci scalar and $Q=R^{\mu\nu}R_{\mu\nu}$).
\par
Let us close this section with a remark concerning how one could coherently make contact with the aforementioned Minkowski limit of the deformed HDA. Since the original Schwarzschild metric already violates Lorentz invariance, in our approach we do not need to consider deformations of the Lorentz symmetry. We would need to be concerned about this issue if we had a Minkowski limit of this metric. However, following the procedures of \cite{BenAchour:2018khr}, we cannot simply place $r_s=0$ because the function $h(K_{\phi})=r_s/t-1$ should be a positive definite function, hence the no-gravity limit needs to be carefully treated, in order to work on the effective spacetime symmetries of this metric description. But, this will be subject for future investigations.

\section{Final remarks}\label{sec-conc}
Based on the recently found black hole solution inside the event horizon from deformations of GR due to quantum gravitational corrections \cite{BenAchour:2018khr}, and on the link between the hypersurface deformation algebra and deformed Poincar\'e algebra in the flat limit \cite{kminklqg}, we connected these two perspectives of the same problem using the, so called, rainbow metrics. In the present case, we found a metric description for the solutions found in \cite{BenAchour:2018khr} based on the relation between the radial triad, the extrinsic curvature and the radial momentum given by $P_{r}=-K_{\phi}/\sqrt{|E^{r}|}$, which is on the very basis of the linearization of the HDA in terms of DSR symmetries. Such metric assumes the form of a rainbow metric, in the sense that it depends on the spacetime coordinates and on the momentum $P_r$.
\par
We analyzed some different realizations of this quantization scheme and realized that a pattern emerged for the general form of the rainbow metric. We have only even functions of the dimensionless quantity $\rho t\, P_r$, that we expanded in a Taylor series and collected the first two terms in table (\ref{table1}).
\par
Important differences with respect to the usual rainbow metric ansatz were found, like the absence of a rainbow function in the line element of the sphere ${\mathbb S}^2$. The presence of a variable, effective Planck length that governs the deformation $\ell_P^{\text {eff}}=\rho t$, which is a novelty in attempts to find rainbow metrics from quantum gravity considerations. And the dependence of the metric on the gravitational and single particle momenta.
\par
We should stress that albeit the effective metrics can already be found from the solution of \cite{BenAchour:2018khr}, we here showed a new ansatz of rainbow metrics, given by Eqs.(\ref{lqg-line-element}) and (\ref{rainbow-ansatz}), inspired by this approach. Alternative formulations of the rainbow gravity initial proposal have been proposed \cite{Girelli:2006fw,Amelino-Camelia:2014rga,Lobo:2016xzq,Lobo:2016lxm,Barcaroli:2015xda,Loret:2014uia,Assaniousssi:2014ota,Lewandowski:2017cvz,Weinfurtner:2008if} and the issue is still under debate. Since the exact form of the semi-classical spacetime description from quantum gravity is not known yet, we should rely on phenomenological possibilities driven by deformation functions, like the HDA approach or rainbow gravity models, for instance.
\par
Another key issue on rainbow gravity and quantum gravity phenomenology in general concerns the deformed trajectories of test particles, i.e., the geodesics of a quantum spacetime. Following an approach similar to ours, some efforts have been pushed forward in \cite{Vakili:2018xws}, and MDRs in flat spacetime have been considered in \cite{kminklqg,mdrlqg2,mrPLB}, which could, in principle, allow us to find trajectories from the Hamilton equations. However, for our purposes it is of pivotal importance to find exterior or near-horizon metric solutions in order to check deviations of the geodesic equations from GR in the direction of confronting our findings with observations and with the near-horizon phenomenology that has been recently developed (see, for instance \cite{Giddings:2016btb}.)
\par
As discussed in \cite{kminklqg,Bojowald:2012ux}, the deformation of the hypersurface algebra induces a deformation of the Poincar\'e algebra. On the other hand, we found that the effective metric description found in \cite{BenAchour:2018khr,Bojowald:2018xxu} resembles rainbow metrics, which are historically related to the DSR program. Therefore, we wonder wether our approach can be useful for discovering an effective metric description of the DSR algebraic formalism, such that trajectories found from deformed Hamilton equations are geodesics and the deformed symmetries are generated by Killing vectors of the metric.
\par
Coherently passing from the ``gravity-on'' to the ``gravity-off'' geometric description, while preserving the aforementioned structures would be an important step towards a ``quantum equivalence principle''. In which the relations between the geometrical quantities in such emergent spacetime is preserved even when considering quantum corrections.
\par
For the future we intend to explore this metric no-gravity limit, in order to find a coherent relativistic metric description of DSR and to better understand the transition from curved to flat metrics in this semiclassical approach.

\vspace{6pt} 


\acknowledgments
The authors thank Suddhasattwa Brahma for reading a preliminary version of the manuscript and his useful comments. This article is based upon work from COST Action CA15117, supported by COST (European Cooperation in Science and Technology. This study was financed in part by the Coordena\c c\~ao de Aperfei\c coamento de Pessoal de N\'ivel Superior - Brasil (CAPES) - Finance Code 001.




\begin{thebibliography}{999}
\bibitem{oriti} Oriti, D. (ed). Approaches to Quantum Gravity. Cambridge University Press: Cambridge U.K., 2009.

\bibitem{smol} Smolin, L. What are we missing in our search for quantum gravity? {\em arXiv} {\bf 2017}, arXiv:1705.09208. 

\bibitem{gacLRR} Amelino-Camelia, G. Quantum-Spacetime Phenomenology. {\em Living Rev. Rel.} {\bf 2013}, {\em 16}, 5, doi:10.12942/lrr-2013-5.

\bibitem{RovelliLRR} Rovelli, C. Loop Quantum Gravity. {\em  Living Rev. Rel.} {\bf 1998}, {\em 1}, 1, doi:10.12942/lrr-1998-1.
            
\bibitem{alvarez} Alvarez, E. In {\em Planck Scale Effects in Astrophysics and Cosmology. Lecture Notes in Physics, vol 669}; Kowalski-Glikman J., Amelino-Camelia G., Eds.; Springer: Berlin, Heidelberg, Germany, 2005; pp. 31-58, 978-3-540-25263-4.

\bibitem{nicolai1} Nicolai, H.; Peeters, K. and Zamaklar, M. Loop quantum gravity: an outside view. {\em Class. Quant. Grav.} {\bf 2005}, {\em 22}, R193, doi:10.1088/0264-9381/22/19/R01. 

\bibitem{nicolai2} Nicolai, H. Quantum Gravity: The View From Particle Physics. In {\em General Relativity, Cosmology and Astrophysics. Fundamental Theories of Physics, vol 177}; Bi\v{c}\'ak J., Ledvinka T., Eds.; Springer: Cham, Switzerland, 2013; pp. 369-387, 978-3-319-06348-5. 

\bibitem{smam} Amelino-Camelia, G. and Smolin, L. Prospects for constraining quantum gravity dispersion with near term observations. {\em Phys.\ Rev. D} {\bf 2009}, {\em 80}, 084017, doi:10.1103/PhysRevD.80.084017. 

\bibitem{liberati1} Carballo-Rubio, R.; Di Filippo, F. and Liberati, S. Minimally modified theories of gravity: a playground for testing the uniqueness of general relativity. {\em JCAP} {\bf 2018}, {\em 1806}, 026, doi:10.1088/1475-7516/2018/06/026. 

\bibitem{liberati2} Carballo-Rubio, R.; Di Filippo, F.; Liberati, S. and Visser, M. Phenomenological aspects of black holes beyond general relativity. {\em arXiv} {\bf 2018}, arXiv:1809.08238. 

\bibitem{mattingly} Mattingly, D. Modern Tests of Lorentz Invariance. {\em Living Rev.\ Rel.} {\bf 2005}, {\em 8}, 5, doi:10.12942/lrr-2005-5. 

\bibitem{dirac} Dirac, P. A. M. An extensible model of the electron. {\em Proc. Roy. Soc. Lond.\ A} {\bf 1962}, {\em 268}, 57, doi:10.1098/rspa.1962.0124.
             
\bibitem{adm}  Arnowitt, R. L.; Deser, S. and Misner, C. W. Republication of: The dynamics of general relativity. {\em Gen. Rel. Grav.} {\bf 2008}, {\em 40}, 1997, doi:10.1007/s10714-008-0661-1.

\bibitem{thibook} Thiemann, T. Modern Canonical Quantum General Relativity. Cambridge University Press, Cambridge U.K.,
2008; 9780511755682.

\bibitem{bojobook} Bojowald, M. Canonical Gravity and Applications: Cosmology, Black Holes, and Quantum Gravity. Cambridge University Press, Cambridge, U.K., 2010; 9780521195751.

\bibitem{CorichiReyes}  Corichi, A. and Reyes, J. D. The gravitational Hamiltonian, first order action, Poincar\'e charges and surface terms. {\em Class. Quant. Grav.} {\bf 2015}, {\em 32}, 195024, doi:10.1088/0264-9381/32/19/195024.

\bibitem{covqg1} Rovelli, C. Covariant Loop Gravity. In {\em Quantum Gravity and Quantum Cosmology. Lecture Notes in Physics, vol 863};  Calcagni G., Papantonopoulos L., Siopsis G., Tsamis N., Eds.; Springer: Berlim, Heidelberg, Germany 2013; pp. 57-66, 978-3-642-33035-3. 

\bibitem{covqg2} Perez, A. Spin foam models for quantum gravity.
{\em Class.\ Quant.\ Grav.} {\bf 2003}, {\em 20}, R43, doi:10.1088/0264-9381/20/6/202. 

\bibitem{alex} Alexandrov, S. and Roche, P. Critical Overview of Loops and Foams.
 {\em Phys.\ Rept.} {\bf 2011}, {\em 506}, 41, doi:10.1016/j.physrep.2011.05.002.

\bibitem{Amb2}  Perez, A. Regularization ambiguities in loop quantum gravity. {\em Phys.\ Rev. D} {\bf 2006}, {\em 73}, 044007, doi:10.1103/PhysRevD.73.044007.

\bibitem{Amb3}  Corichi, A. and Singh, P. Is loop quantization in cosmology unique? {\em Phys.\ Rev. D} {\bf 2008}, {\em 78}, 024034, doi:10.1103/PhysRevD.78.024034.

\bibitem{CaitellMielcBarr} Cailleteau, T.; Mielczarek, J.; Barrau, A. and Grain, J. Anomaly-free scalar perturbations with holonomy corrections in loop quantum cosmology. {\em Class. Quant. Grav.} {\bf 2012}, {\em 29}, 095010, doi:10.1088/0264-9381/29/9/095010.

\bibitem{AshtLewMarMouThie} Ashtekar, A.; Lewandowski, J.; Marolf, D.; Mourao, J. and Thiemann, T. Quantization of diffeomorphism invariant theories of connections with local degrees of freedom. {\em J. Math. Phys.} {\bf 1995}, {\em 36}, 6456, doi:10.1063/1.531252.

\bibitem{bojopail2} Bojowald, M. and Paily, G. M. Deformed general relativity and effective actions from loop quantum gravity. {\em Phys. Rev. D} {\bf 2012}, {\em 86}, 104018, doi:10.1103/PhysRevD.86.104018.

\bibitem{bojoHossKag} Bojowald, M.; Hossain, G.M.; Kagan, M. and Shankaranarayanan, S. Anomaly freedom in perturbative loop quantum gravity. {\em Phys. Rev. D} {\bf 2008} {\em 78}, 063547, doi:10.1103/PhysRevD.78.063547.

\bibitem{Thiemann2} Thiemann. T. The Phoenix Project: master constraint programme for loop quantum gravity. {\em Class. Quant. Grav.} {\bf 2006}, {\em 23}, 2211, doi:10.1088/0264-9381/23/7/002.

\bibitem{CuttSakell} Cuttell, P. D. and Sakellariadou, M. Fourth order deformed general relativity.
 {\em Phys. Rev. D} {\bf 2014}, {\em 90}, 104026, doi:10.1103/PhysRevD.90.104026.

\bibitem{BojoBraRey} Bojowald, M.; Brahma, S. and Reyes, J. D. Covariance in models of loop quantum gravity: Spherical symmetry. {\em Phys. Rev.\ D} {\bf 2015}, {\em 92}, 045043, doi:10.1103/PhysRevD.92.045043.
            
\bibitem{B1} Bojowald, M. and Brahma, S. Covariance in models of loop quantum gravity: Gowdy systems.
 {\em Phys.\ Rev. D} {\bf 2015}, {\em 92}, 065002, doi:10.1103/PhysRevD.92.065002. 

\bibitem{B2} Bojowald, M.; Brahma, S.; Buyukcam,U. and D'Ambrosio, F. Hypersurface-deformation algebroids and effective spacetime models. {\em Phys.\ Rev. D} {\bf 2016}, {\em 94}, 104032, doi:10.1103/PhysRevD.94.104032. 

\bibitem{B3} Wu, J. P.; Bojowald, M. and Ma, Y. Anomaly freedom in perturbative models of Euclidean loop quantum gravity. {\em arXiv} {\bf 2018}, arXiv:1809.04465. 

\bibitem{calcagni} Calcagni, G. and Ronco, M. Deformed symmetries in noncommutative and multifractional spacetimes, {\em Phys.\ Rev. D} {\bf 2017}, {\em 95}, 045001, doi:10.1103/PhysRevD.95.045001. 

\bibitem{moyaldiff} Bojowald, M.; Brahma, S.; Buyukcam, U. and Ronco, M. Extending general covariance: Moyal-type noncommutative manifolds. {\em Phys.\ Rev. D} {\bf 2018}, {\em 98}, 026031, doi:10.1103/PhysRevD.98.026031. 

\bibitem{dsr1} Amelino-Camelia, G. Relativity in space-times with short-distance structure governed by an observer-independent (Planckian) length scale. {\em Int. J. Mod. Phys. D} {\bf 2002}, {\em 11}, 35, doi:10.1142/S0218271802001330.

\bibitem{dsr2} Magueijo, J. and Smolin, L. Lorentz invariance with an invariant energy scale. {\em Phys. Rev. Lett.} {\bf 2002}, {\em 88}, 190, doi:10.1103/PhysRevLett.88.190403.

 \bibitem{Magueijo:2002am} 
  Magueijo, J. and Smolin, L.
 Generalized Lorentz invariance with an invariant energy scale.
  {\em Phys.\ Rev.\ D} {\em 2003}, {\bf 67}, 044017, doi:10.1103/PhysRevD.67.044017.

\bibitem{dsr3} Amelino-Camelia, G. Limits on the Measurability of Space-time Distances in (the Semi-classical Approximation of) Quantum Gravity. {\em Mod. Phys. Lett. A} {\bf 1994}, {\em 9}, 3415, doi:10.1142/S0217732394003245.

\bibitem{dsr4} Amelino-Camelia, G. Testable scenario for relativity with minimum length. {\em Phys. Lett. B} {\bf 2001}, {\em 510}, 255, doi:10.1016/S0370-2693(01)00506-8.

\bibitem{majRue} Majid, S. and Ruegg, H. Bicrossproduct structure of $\kappa$-Poincar\'e group and non-commutative geometry. {\em Phys. Lett. B} {\bf 1994}, {\em 334}, 348, doi:10.1016/0370-2693(94)90699-8.

\bibitem{lukRue} Lukierski, J.; Ruegg, H. and Zakrzewski, W. J. Classical and Quantum Mechanics of Free $\kappa$-Relativistic Systems. {\em Ann. Phys.} {\bf 1995}, {\em 243}, 90, doi:10.1006/aphy.1995.1092. 

\bibitem{kminklqg} Amelino-Camelia, G.; da Silva, M. M.; Ronco, M.; Cesarini, L. and Lecian, O. M. Spacetime-noncommutativity regime of loop quantum gravity {\em Phys.\ Rev. D} {\bf 2017}, {\em 95}, 024028, doi:10.1103/PhysRevD.95.024028. 
  
\bibitem{mdrlqg1} Mielczarek, J. Loop-deformed Poincar\'e algebra. {\em EPL} {\bf 2014}, {\bf 108}, 40003, doi:10.1209/0295-5075/108/40003. 

\bibitem{mdrlqg2} Brahma, S.; Ronco, M.; Amelino-Camelia, G. and Marcian\`o, A. Linking loop quantum gravity quantization ambiguities with phenomenology. {\em Phys.\ Rev. D} {\bf 2017}, {\em 95}, 044005, doi:10.1103/PhysRevD.95.044005 . 

\bibitem{lqgdim1} Ronco, M. On the UV dimensions of Loop Quantum Gravity. {\em Adv.\ High Energy Phys.} {\bf 2016}, {\em 2016}, 9897051, doi:10.1155/2016/9897051 . 

\bibitem{lqgdim2} Mielczarek, J. and Trześniewski, T. Spectral dimension with deformed spacetime signature. {\em Phys.\ Rev. D} {\bf 2017}, {\em 96}, 024012, doi:10.1103/PhysRevD.96.024012. 

\bibitem{mrPLB} Brahma, S. and Ronco, M. Constraining the loop quantum gravity parameter space from phenomenology. {\em Phys.\ Lett. B} {\bf 2018}, {\em 778}, 184, doi:10.1016/j.physletb.2018.01.023. 

\bibitem{Bojowald:2018xxu} 
  Bojowald, M.; Brahma, S. and Yeom, D. h.
 Effective line elements and black-hole models in canonical loop quantum gravity. {\em Phys.\ Rev. D} {\bf 2018}, {\em 98}, 046015, doi:10.1103/PhysRevD.98.046015.
  
\bibitem{BenAchour:2018khr} 
  Ben Achour, J.; Lamy, F.; Liu, H. and Noui, K.
  Polymer Schwarzschild black hole: An effective metric.
  {\em EPL} {\bf 2018}, {\em 123}, 20006, doi:10.1209/0295-5075/123/20006.

\bibitem{Bianchi:2009ky} 
  Bianchi, E.; Magliaro, E. and Perini, C. Coherent spin-networks. {\em Phys.\ Rev.\ D} {\bf 2010}, {\em 82}, 024012, doi:10.1103/PhysRevD.82.024012.

    
\bibitem{Magueijo:2002xx} 
  Magueijo, J. and Smolin, L.
Gravity's rainbow.
  {\em Class.\ Quant.\ Grav.} {\bf 2004},  {\em 21}, 1725, doi:10.1088/0264-9381/21/7/001.

\bibitem{Galan:2004st} 
  Galan, P. and Mena Marugan, G. A.
Quantum time uncertainty in a gravity's rainbow formalism.
  {\em Phys.\ Rev.\ D} {\bf 2004}, {\em 70}, 124003,
 doi:10.1103/PhysRevD.70.124003.

\bibitem{Ali:2014xqa} 
  Ali, A. F. Black hole remnant from gravity's rainbow.
  {\em Phys.\ Rev.\ D} {\bf 2014}, {\em 89}, 104040,
  doi:10.1103/PhysRevD.89.104040.
  
\bibitem{Heydarzade:2017rpb} 
  Heydarzade, Y.; Rudra, P.; Darabi, F.; Ali, A. F. and Faizal, M.
  Vaidya spacetime in massive gravity's rainbow.
  {\em Phys.\ Lett.\ B} {\bf 2017}, {\em 774}, 46,
  doi:10.1016/j.physletb.2017.09.049
  
\bibitem{Barrau} Barrau, A.; Bojowald, M.; Calcagni, G.; Grain, J. and Kagan, M. Anomaly-free cosmological perturbations in effective canonical quantum gravity. {\em JCAP} {\bf 2015}, {\em 05}, 1505, doi:10.1088/1475-7516/2015/05/051.

\bibitem{thieTr} Giesel, K. and Thiemann, T. Consistency check on volume and triad operator quantization in loop quantum gravity: I. {\em Class.\ Quant.\ Grav.} {\bf 2006}, {\em 23}, 5667, doi:10.1088/0264-9381/23/18/011. 

\bibitem{bra1} Bojowald, M. and Brahma, S. Signature change in two-dimensional black-hole models of loop quantum gravity. {\em Phys.\ Rev. D} {\bf 2018}, {\em 98}, 026012, doi:10.1103/PhysRevD.98.026012. 

\bibitem{bra2} Brahma, S. Spherically symmetric canonical quantum gravity. {\em Phys.\ Rev. D} {\bf 2015}, {\em 91}, 124003, doi:10.1103/PhysRevD.91.124003.  

\bibitem{SSnonlocalhol} Bojowald, M.; Paily, G. M. and Reyes, J. D. Discreteness corrections and higher spatial derivatives in effective canonical quantum gravity. {\em Phys.\ Rev. D} {\bf 2014}, {\em 90}, 025025, doi:10.1103/PhysRevD.90.025025.

\bibitem{Varadarajan} Varadarajan, M. The diffeomorphism constraint operator in loop quantum gravity. {\em J.\ Phys.\ Conf.\ Ser.} {\bf 2012}, {\em 360}, 012009, doi:10.1088/1742-6596/360/1/012009. 

\bibitem{olmedo} Olmedo, J. Brief Review on Black Hole Loop Quantization. {\em Universe} {\bf 2016}, {\em 2}, 12, doi:10.3390/universe2020012. 

\bibitem{ed} Wilson-Ewing, E. Holonomy corrections in the effective equations for scalar mode perturbations in loop quantum cosmology. {\em Class.\ Quant.\ Grav.} {\bf 2012}, {\em 29}, 085005, doi:10.1088/0264-9381/29/8/085005.

\bibitem{fra} Cailleteau, T.; Barrau, A.; Grain, J. and Vidotto, F. Consistency of holonomy-corrected scalar, vector, and tensor perturbations in loop quantum cosmology. {\em Phys.\ Rev. D} {\bf 2012}, {\em 86}, 087301, doi:10.1103/PhysRevD.86.087301. 

\bibitem{BY} Brown, J. D.; York, J. W. Quasilocal energy and conserved charges derived from the gravitational action. {\em Phys.\ Rev. D} {\bf 1993}, {\em 47}, 1407, doi:10.1103/PhysRevD.47.1407. 

\bibitem{born}
Born, M. A Suggestion for Unifying Quantum Theory and Relativity. {\em Proc. R. Soc. Lond. A} {\bf 1938}, {\em 165}, 291, doi:10.1098/rspa.1938.0060.

\bibitem{Bojowald:2012ux} 
  Bojowald, M. and Paily, G. M.
 Deformed General Relativity.
  {\em Phys.\ Rev.\ D} {\bf 2013}, {\em 87}, 044044, doi:10.1103/PhysRevD.87.044044.

\bibitem{su11}  Ben Achour, J.; Mouchet, A. and Noui, K. Analytic Continuation of Black Hole Entropy in Loop Quantum Gravity. {\em JHEP} {\bf 2015}, {\em 1506}, 145, doi:10.1007/JHEP06(2015)145.

\bibitem{Achour:2014rja} 
  Ben Achour, J.; Grain, J. and Noui, K.
  Loop Quantum Cosmology with Complex Ashtekar Variables.
  {\em Class.\ Quant.\ Grav.} {\bf 2015}, {\em 32}, 025011, doi:10.1088/0264-9381/32/2/025011.


\bibitem{cdt} Ambjørn, J.; Jurkiewicz, J. and Loll, R. The Spectral Dimension of the Universe is Scale Dependent. {\em Phys.\ Rev.\ Lett.} {\bf 2005}, {\em 95}, 171301, doi:10.1103/PhysRevLett.95.171301. 

\bibitem{as} Niedermaier, M. and Reuter, M. The Asymptotic Safety Scenario in Quantum Gravity. {\em Living Rev.\ Rel.} {\bf 2006}, {\em 9}, 5, doi:10.12942/lrr-2006-5. 

\bibitem{calcagni2} Calcagni, G. Multifractional theories: an unconventional review. {\em JHEP} {\bf 2017}, {\em 1706}, 020, doi:10.1007/JHEP06(2017)020.  

\bibitem{mrPLB2} Amelino-Camelia, G.; Calcagni, G. and Ronco, M. Imprint of quantum gravity in the dimension and fabric of spacetime. {\em Phys.\ Lett. B} {\bf 2017}, {\em 774}, 630,doi:10.1016/j.physletb.2017.10.032. 

\bibitem{long} Calcagni, G. and Ronco, M.; 	
Dimensional flow and fuzziness in quantum gravity: emergence of stochastic spacetime. 
{\em Nucl.\ Phys. B} {\bf 2017}, {\em 923}, 144, doi:10.1016/j.nuclphysb.2017.07.016. 

  \bibitem{Olmo:2011sw} 
  Olmo, G.J.
Palatini Actions and Quantum Gravity Phenomenology.
  {\em JCAP} {\bf 2011}, {\em 1110}, 018, doi:10.1088/1475-7516/2011/10/018.

\bibitem{Girelli:2006fw} 
  Girelli, F.; Liberati, S. and Sindoni, L.
  ``Planck-scale modified dispersion relations and Finsler geometry.
  {\em Phys.\ Rev.\ D} {\bf 2007}, {\em 75}, 064015, doi:10.1103/PhysRevD.75.064015.

\bibitem{Amelino-Camelia:2014rga} 
  Amelino-Camelia, G.; Barcaroli, L.; Gubitosi, G.; Liberati, S. and Loret, N.
  Realization of doubly special relativistic symmetries in Finsler geometries.
  {\em Phys.\ Rev.\ D} {\bf 2014}, {\em 90}, 125030, doi:10.1103/PhysRevD.90.125030.

\bibitem{Lobo:2016xzq} 
 Lobo, I. P.; Loret, N. and Nettel, F.
 Investigation of Finsler geometry as a generalization to curved spacetime of Planck-scale-deformed relativity in the de Sitter case.
  {\em Phys.\ Rev.\ D} {\bf 2017}, {\em 95}, 046015, doi:10.1103/PhysRevD.95.046015.
 
\bibitem{Lobo:2016lxm} 
  Lobo, I. P.; Loret, N. and Nettel, F.
Rainbows without unicorns: Metric structures in theories with Modified Dispersion Relations.
  {\em Eur.\ Phys.\ J.\ C} {\bf 2017}, {\em 77}, 451, doi:10.1140/epjc/s10052-017-5017-0.
 
\bibitem{Barcaroli:2015xda} 
  Barcaroli, L.; Brunkhorst, L. K.; Gubitosi, G.; Loret, N. and Pfeifer, C.
Hamilton geometry: Phase space geometry from modified dispersion relations.
  {\em Phys.\ Rev.\ D} {\bf 2015}, {\em 92}, 084053, doi:10.1103/PhysRevD.92.084053.

\bibitem{Loret:2014uia} 
  Loret, N.
Exploring special relative locality with de Sitter momentum-space.
  {\em Phys.\ Rev.\ D} {\bf 2014}, {\em 90}, 124013, doi:10.1103/PhysRevD.90.124013.

\bibitem{Assaniousssi:2014ota} 
  Assanioussi, M.; Dapor, A. and Lewandowski, J. Rainbow metric from quantum gravity.
  {\em Phys.\ Lett.\ B} {\bf 2015}, {\em 751}, 302,
  doi:10.1016/j.physletb.2015.10.043.

\bibitem{Lewandowski:2017cvz} 
  Lewandowski, J.; Nouri-Zonoz, M.; Parvizi, A. and Tavakoli, Y. Quantum theory of electromagnetic fields in a cosmological quantum spacetime.
  {\em Phys.\ Rev.\ D} {\bf 2017}, {\em 96}, 106007, doi:10.1103/PhysRevD.96.106007.
  
\bibitem{Weinfurtner:2008if} 
  Weinfurtner, S.; Jain, P.; Visser, M. and Gardiner, C. W. Cosmological particle production in emergent rainbow spacetimes.
  {\em Class.\ Quant.\ Grav.} {\bf 2009}, {\em 26}, 065012, doi:10.1088/0264-9381/26/6/065012

\bibitem{Vakili:2018xws} 
  Vakili, B. Classical polymerization of the Schwarzschild metric. 
  {\em Adv.\ High Energy Phys.} {\bf 2018}, {\em 2018}, 3610543, doi:10.1155/2018/3610543.

\bibitem{Giddings:2016btb} 
  Giddings, S. B. and Psaltis D. Event Horizon Telescope Observations as Probes for Quantum Structure of Astrophysical Black Holes.
  {\em Phys.\ Rev.\ D} {\bf 2018}, {\em 97}, 084035, doi:10.1103/PhysRevD.97.084035.

\end{thebibliography}



\end{document}